\documentclass[]{rspublic}
\usepackage{mathrsfs,amsfonts,graphics,cancel,color}
\usepackage[english]{babel}  
\usepackage{natbib}

\def\bra#1{\langle{#1}|}
\def\ket#1{|{#1}\rangle}
\def\ketbra#1{|{#1}\rangle\langle{#1}|}
\def\braket#1{\langle{#1}\rangle}

{\catcode`\|=\active 
  \gdef\Braket#1{\begingroup
\mathcode`\|32768\let|\BraVert\left<{#1}\right>\endgroup}}
\def\BraVert{\egroup\,\mid\,\bgroup}

\definecolor{Blue}{rgb}{0,0,1}
\definecolor{Red}{rgb}{0,0,0}
\definecolor{Green}{rgb}{0,1,0}
\definecolor{Purp}{rgb}{.2,0,.2}
\definecolor{white}{rgb}{1,1,1}

\newcommand{\tr}{\mathop{\mathrm{tr}}\nolimits}
\newcommand{\Oc}{\mathcal{J}}

\begin{document}

\date{\today}
\title{Classical to quantum in large number limit}
\author[K. Modi, R. Fazio, S. Pascazio, V. Vedral, and K. Yuasa]{Kavan Modi$^{1,2}$, Rosario Fazio$^{3,2}$, Saverio Pascazio$^{4,5}$, Vlatko Vedral$^{1,2,6}$, and Kazuya Yuasa$^7$}
\affiliation{
$^1$Department of Physics, University of Oxford, Clarendon Laboratory, Oxford UK\\
$^2$Centre for Quantum Technologies, National University of Singapore, Singapore\\
$^3$NEST, Scuola Normale Superiore and Istituto Nanoscienze -- CNR, Pisa, Italy\\
$^4$Dipartimento di Fisica and MECENAS, Universit\`a di Bari, Bari, Italy\\
$^5$INFN, Sezione di Bari, Bari, Italy\\
$^6$Department of Physics, National University of Singapore, Singapore\\
$^7$Waseda Institute for Advanced Study, Waseda University, Tokyo 169-8050, Japan}

\maketitle
\begin{abstract}
{quantumness witness, purity, large number limit}
We construct a quantumness witness following the work of Alicki and van Ryn (AvR) in ``A simple test of quantumness for a single system" [\textit{J. Phys.\ A: Math.\ Theor.}, \textbf{41}, 062001 (2008)]. The AvR test is designed to detect quantumness. We reformulate the AvR test by defining it for quantum states rather than for observables. This allows us to identify the necessary quantities and resources to detect quantumness for any given system. The first quantity turns out to be the purity of the system. When applying the witness to a system with even moderate mixedness the protocol is unable to reveal any quantumness. We then show that having many copies of the system leads the witness to reveal quantumness. This seems contrary to the Bohr correspondence, which asserts that in the large number limit quantum systems become classical, while the witness shows quantumness when several non-quantum systems, as determined by the witness, are considered together. However, the resources required to detect the quantumness increase dramatically with the number of systems. We apply the quantumness witness for systems that are highly mixed but in the large number limit, which resembles nuclear magnetic resonance (NMR) systems. We make several conclusion about detecting quantumness in NMR-like systems.

\end{abstract}

%%%%%%%%%%%%%%%%%%%%%%%%%%%%%%%%%%%%%%%%%%%%%%%%%%%%%%%%%%%%%%%
%%%%%%%%%%%%%%%%%%%%%%%%%%%%%%%%%%%%%%%%%%%%%%%%%%%%%%%%%%%%%%%
\section{Introduction}
%%%%%%%%%%%%%%%%%%%%%%%%%%%%%%%%%%%%%%%%%%%%%%%%%%%%%%%%%%%%%%%
%%%%%%%%%%%%%%%%%%%%%%%%%%%%%%%%%%%%%%%%%%%%%%%%%%%%%%%%%%%%%%%

A fundamental quest of modern physics is to characterize the crossover between the quantum world and the classical world. There are many different approaches to answer this question, yet no single approach captures the whole breath of physics. It is clear that for some systems classical physics arises from quantum physics in the large number limit. This is the Bohr correspondence principle. Yet it is not fully understood how the many-particle limit gives rise to classical physics, and how much of quantum physics still remains. Simply put there are many operational tasks that give sufficient conditions for quantumness of a system. Yet no single system may satisfy all of these conditions. Therefore it is desirable to have a broad notion of quantumness that encompasses all (or many) other notions of quantumness.

One answer in this quest that compels many is given by the violations of Bell's inequalities. Bell's theorem gives a recipe to test for strictly quantum features of two spatially separated systems, namely entanglement and nonlocality. More often than not we deal with systems composed of many subsystems that are not spatially distributed, and therefore nonlocality may not be the essential quantum feature of the system of interest. Defining quantumness of isolated systems remains a challenge.

The papers of Alicki, van Ryn (AvR), and later with Piani~(\cite{AVR,APVR}) give a test for quantumness of a single system. This test does not necessarily encompass all aspects of quantumness of a system~(\cite{Zukowski,Alicki}), nevertheless it offers a glimpse at the classical--quantum divide. On the other hand, it should be noted that all entanglement witnesses are also quantumness witnesses, and an important class of entanglement witnesses are actually quantumness witnesses of the AvR type~(\cite{FFPVY}). In this article we propose a quantumness witness based on the AvR test. Our witness allows for resource accounting, in terms of qubits and controlled-\textsc{shift} gates, to detect quantumness.

%%%%%%%%%%%%%%%%%%%%%%%%%%%%%%%%%%%%%%%%%%%%%%%%%%%%%%%%%%%%%%%
%%%%%%%%%%%%%%%%%%%%%%%%%%%%%%%%%%%%%%%%%%%%%%%%%%%%%%%%%%%%%%%
\section{Quantumness test}
%%%%%%%%%%%%%%%%%%%%%%%%%%%%%%%%%%%%%%%%%%%%%%%%%%%%%%%%%%%%%%%
%%%%%%%%%%%%%%%%%%%%%%%%%%%%%%%%%%%%%%%%%%%%%%%%%%%%%%%%%%%%%%%

The quantumness test due to AvR is based on a mathematical theorem. AvR give a set of simple criteria for classicality; when these criteria are violated by a system, the system is said to be quantum. The criteria are based on observables of the system. The test presented in this article is based on the same mathematical principles, but we work with states of the system rather than the observables of the system. We show that working with states gives new insight into this notion of quantumness. Namely, (\cite{AVR,APVR}) show that for any quantum state there exists a quantumness witness as defined by AvR. This is also true in the state approach, but are able to quantify the resources, number states needed, to witness quantumness. We begin by reviewing the quantumness test due to AvR and then derive our witness.

%%%%%%%%%%%%%%%%%%%%%%%%%%%%%%%%%%%%%%%%%%%%%%%%%%%%%%%%%%%%%%%
\subsection{Observable approach}
%%%%%%%%%%%%%%%%%%%%%%%%%%%%%%%%%%%%%%%%%%%%%%%%%%%%%%%%%%%%%%%

To test the quantumness of a single system, consisting of many particles, AvR note the following mathematical theorem.

{\bf Theorem 1.} {\sl Given a $C^*$--algebra $\mathcal{A}$, the following three statements are equivalent: 
\begin{itemize}
\item[i.] $\mathcal{A}$ is commutative, i.e. for any pair $X,Z\in\mathcal{A}$, $[X,Z]=XZ-ZX=0.$
%\begin{gather} \label{classicalB} \end{gather}
\item[ii.] For any pair $X,  Z\in\mathcal{A}$ with $0 \le X \le Z$ implies $X^2 \le Z^2.$
%\begin{gather} X^2 \le Y^2. \label{classicalC} \end{gather}
\item[iii.] For any pair $X,  Z \in \mathcal{A}$ with $X \geq 0$ and $ Z \geq 0$ implies that the operator corresponding to the anticommutator is also positive:
\begin{gather}
\{X,Z\}= {XZ + ZX} \geq 0 . \label{classicalA}
\end{gather}
\end{itemize}}
The proof is given in~(\cite{AVR, APVR}) and the reference~(\cite{Dix,Kad,Oga}). Above $A$ is non-negative if it can be written in the form $A=B^\dagger B$ for some $B$, and $C \geq D$ means that $C-D$ is non-negative.
%$0 \ge X \ge Z$ means the expectation values of the observables $Z$ is greater than $X$ and positive.

AvR make use of statements (i) and (ii) to define a quantumness witness. Theorem 1 states that if $\mathcal{A}$ is noncommutative there exist two noncommuting positive observables satisfying $X \ge Z$ for all states, but there also exists a state $\phi$ such that $\braket{X^2}_\phi < \braket{Z^2}_\phi$. This is a violation of the classical principle. In~(\cite{APVR}) the authors show that for every quantum state, excluding the maximally mixed state, there exists an observable of statement (iii) type. This quantumness witness was experimentally tested in~(\cite{brida08, bridagisin}).

The AvR test can be thought of as a test for complementarity. Observables in classical physics have simultaneous values, but not in quantum mechanics. Any two observables that do not commute do not have simultaneous values and are said to be complementary. For instance position and momentum are complementary observable in quantum theory and do not have simultaneous values.  As Bell's inequalities test for local realism, AvR criteria test for whether two observable have simultaneous values for a given quantum system in a specific state.

Complementary observables in quantum theory are similar to states of a quantum system that cannot be fully discriminated. For instance if we are given a state, $\ket{0}$ or $\ket{+}$, we will not be able to determine, with full confidence, which state was given.
Another way to think about this is that two states that cannot be discriminated are eigenstates of observables that do not commute. Based on that intuition we now construct a quantumness witness using statement (iii) of theorem 1 with quantum states rather than the observable approach taken in~(\cite{AVR, APVR}). We will show that this gives us a tool to quantify the necessary resource for witnessing quantumness.

%%%%%%%%%%%%%%%%%%%%%%%%%%%%%%%%%%%%%%%%%%%%%%%%%%%%%%%%%%%%%%%
\subsection{State approach}
%%%%%%%%%%%%%%%%%%%%%%%%%%%%%%%%%%%%%%%%%%%%%%%%%%%%%%%%%%%%%%%

Let the Hilbert space of the quantum system be $\mathcal{H} \simeq \mathbb{C}^d$. We begin by noting that any Hermitian element $X\in\mathcal{A}$ is a quantum observable. Any Hermitian observable can be expressed
\begin{gather}
X= x_0 \mathbb{I} + \sum_i x_i \sigma_i,
\end{gather}
where $\mathbb{I}$ is the identity matrix and $\sigma_i$ are Hermitian basis matrices (traceless generators of the $SU(d)$ group). Hermiticity restricts $x_0, x_i$ to be real and if the observable is positive then $x_0 > 0$ and $x_i$ as a function of $x_0$. 

If the observable is positive then it is related to a quantum state in a simple manner. A density operator, that represents the state of a $d$--dimensional quantum system, is similarly written as
\begin{gather}
\rho= \frac{1}{d} \left(\mathbb{I} + \sum_i r_i \sigma_i \right),
\end{gather}
where again $r_i$ will have some restrictions due to positivity of $\rho$ (\cite{Kimura}). The difference between an observable and a density operator is simply the normalization:
\begin{gather}
\tr[X]=x_0 , \quad \tr[\rho]=1.
\end{gather}
There is a clear mapping between density operators and positive observables:
\begin{gather}
X = (x_0d)\rho_X \quad \Longleftrightarrow \quad \rho_X = \frac{1}{d} \left(\mathbb{I} + \sum_i \frac{x_i}{x_0} \sigma_i \right).
\end{gather}

The normalization of a matrix does not alter its positivity. That is, if the operator $XZ+ZX$ is positive then so is the corresponding density operator $\rho_X \rho_Z + \rho_Z \rho_X$. The duality between states and observables gives us a quantumness witness using statement (iii) of theorem 1 for Eq.~(\ref{classicalA}).

{\bf Definition 1.} {\sl For every pair of states $\rho_X$ and $\rho_Z$ a system can be prepared into, let $\Oc$ be the operator of the anticommutator:
\begin{gather}
\label{witness1}
\Oc=\{ \rho_X , \rho_Z \} = \rho_X \rho_Z + \rho_Z \rho_X.
\end{gather}
The system is quantum if it can be prepared into two states $\rho_X$ and $\rho_Z$ such that the corresponding anticommutator operator $\Oc \ngeqslant 0$.}

%Additionally we may want to define the degree of quantumness based on this method. {\bf Definition 2.} {\sl Let $\{\lambda_i\}$ be the eigenvalues of $\Oc$, which we can split into the negative and positive sets: $\{\lambda_{i^-},\lambda_{i^+}\}$. Let us define the degree of quantumness as $\mathcal{Q}=\sum_{i^-}\lambda_{i^-}$. If $\lambda_{\min} \ge 0$, then the system is classical.} The definition of degree of quantumness may not be taken seriously, as it is somewhat arbitrary. However, it measures how negative the operator $\Oc$ can get and whether this negativity is measurable. We will show that for some systems the negativity is so small that it will take tremendously large measuring apparatus to sense it. This is precisely what we observe in our classical world, that it takes a lot of work to measure quantumness.

%%%%%%%%%%%%%%%%%%%%%%%%%%%%%%%%%%%%%%%%%%%%%%%%%%%%%%%%%%%%%%%
%%%%%%%%%%%%%%%%%%%%%%%%%%%%%%%%%%%%%%%%%%%%%%%%%%%%%%%%%%%%%%%
\section{Observing quantumness}\label{OQ}
%%%%%%%%%%%%%%%%%%%%%%%%%%%%%%%%%%%%%%%%%%%%%%%%%%%%%%%%%%%%%%%
%%%%%%%%%%%%%%%%%%%%%%%%%%%%%%%%%%%%%%%%%%%%%%%%%%%%%%%%%%%%%%%

Our quantumness witness is experimentally testable. Note that 
\begin{gather}
\tr[\Oc] = \tr[\rho_X \rho_Z+\rho_Z \rho_X] = 2 \tr[\rho_X \rho_Z]= \sum_i \lambda_i,
\end{gather}
where $\lambda_i$ are the eigenvalues of $\Oc$. We are interested in the eigenvalues of $\Oc$, not the sum of the eigenvalues. Since there are $d$ eigenvalues and if we know the values of $\tr[\Oc^l]=\sum_i \lambda_i^l$ for all values of $1 \le l \le d$, then we can algebraically solve for $\{\lambda_i\}$. To find the eigenvalues of the $\Oc$ we can measure 
\begin{gather}
j_l=\tr[ (\rho_X \rho_Z)^l].
\end{gather}
The trace of product of two density operator is a real-positive number: $\tr[ \rho_X \rho_Z ]=\sum_{xz} r_x r_z|{\braket{x|z}}|^2$, where $r_x$ $(r_z)$ and $\ket{x}$ $(\ket{z})$ are the eigenvalues and eigenvectors of $\rho_X$ $(\rho_Z)$, respectively.

To measure the product of two operators in an experimental setting we make use of the controlled-\textsc{shift} operator~(\cite{PhysRevA.64.052311,PhysRevLett.88.217901,vlatkoqfi}). The \textsc{shift} operator's action is defined as: 
\begin{gather}
S\ket{\psi_1,\psi_2,\dots,\psi_{l-1},\psi_{l}}=
\ket{\psi_l,\psi_1,\psi_2,\dots,\psi_{l-1}}.
\end{gather}
The trace of the \textsc{shift} operator's action from one side only yields
\begin{gather}
\tr[S(\rho_1 \otimes \rho_2 \otimes \dots \otimes \rho_l)] = \tr[\rho_1 \rho_2 \cdots\rho_{l}].
\end{gather}
Notice the tensor product on the l.h.s.\ and the ordinary matrix product on the r.h.s.

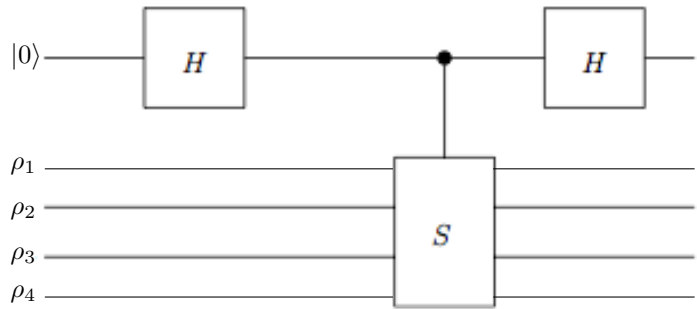
\begin{figure}[t]
\begin{center}
\resizebox{8.76 cm}{4.35 cm}{\includegraphics{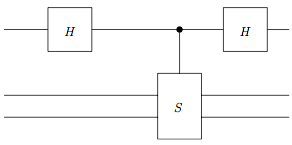}}
\begin{picture}(2.2,3)
\setlength{\unitlength}{10cm}
\put(-0.92,.34){$\ket{0}$} 
\put(-0.92,.2){$\rho_1$}
\put(-0.92,.14){$\rho_2$}
\put(-0.92,.08){$\rho_3$}
\put(-0.92,.03){$\rho_4$}
\put(-0.875,.2){\line(1,0){.458}}
\put(-0.875,.03){\line(1,0){.458}}  
\put(-0.285,.2){\line(1,0){.265}}
\put(-0.285,.03){\line(1,0){.265}}  
\end{picture}
\end{center}
\caption{\label{swap} $H$ is the Hadamard gate, $S$ is the \textsc{shift} gate. The expectation value of the control qubit (top qubit) in $z$--basis is the trace of the product $\tr[\rho_1\rho_2\rho_3\rho_4]$.}
\end{figure}

Using a control qubit and implementing the controlled-\textsc{shift} operator, $\mathcal{C}_{S} = \ketbra{0} \otimes \mathbb{I} + \ketbra{1} \otimes {S}$, we can measure the trace of the product of any number of operators, see Fig.~\ref{swap} for a circuit representation. Measuring the control qubit in $z$--basis will give the product of the density operators. However to measure the $l$th power we will need $l$ copies of the density operator and a suitable \textsc{shift} operation. These resources may not be trivially available and therefore witnessing the quantumness may prove to be difficult.

%%%%%%%%%%%%%%%%%%%%%%%%
%%%%%%%%%%%%%%%%%%%%%%%%

%%%%%%%%%%%%%%%%%%%%%%%%%%%%%%%%%%%%%%%%%%%%%%%%%%%%%%%%%%%%%%%
%%%%%%%%%%%%%%%%%%%%%%%%%%%%%%%%%%%%%%%%%%%%%%%%%%%%%%%%%%%%%%%
\section{Examples of quantumness witness}
%%%%%%%%%%%%%%%%%%%%%%%%%%%%%%%%%%%%%%%%%%%%%%%%%%%%%%%%%%%%%%%
%%%%%%%%%%%%%%%%%%%%%%%%%%%%%%%%%%%%%%%%%%%%%%%%%%%%%%%%%%%%%%%

Now we put this quantumness witness to work with a few examples. The example will illustrate that our witness identifies the crossover between the quantum and classical theories. We consider a system that can be prepared in a set of commuting states, pure states, and mixed states. The first example identifies a ``truly classical'' system, while the second gives a ``truly quantum'' system, and in the final example quantum and classical worlds are blended together.

%%%%%%%%%%%%%%%%%%%%%%%%%%%%%%%%%%%%%%%%%%%%%%%%%%%%%%%%%%%%%%%
\subsection{Commuting states}
%%%%%%%%%%%%%%%%%%%%%%%%%%%%%%%%%%%%%%%%%%%%%%%%%%%%%%%%%%%%%%%

Suppose a system can be prepared into states that all commute with each other and in this sense are viewed as ``classical" by the anticommutator witness in Eq.~(\ref{witness1}). The density operators of such states then can be simultaneously diagonalized. Therefore $\Oc$ is also diagonal in the same basis as the states and its eigenvalues are twice the products of eigenvalues of $\rho_X$ and $\rho_Y$, which are positive numbers. It is then clear that $\Oc \ge 0$ when all states of the system commute, which is in full agreement with  theorem 1.

%%%%%%%%%%%%%%%%%%%%%%%%%%%%%%%%%%%%%%%%%%%%%%%%%%%%%%%%%%%%%%%
\subsection{Pure states}
%%%%%%%%%%%%%%%%%%%%%%%%%%%%%%%%%%%%%%%%%%%%%%%%%%%%%%%%%%%%%%%

Pure states show highly quantum behavior. One can show that an operator of the form
\begin{gather}
\Oc= \braket{\psi|\phi}|\psi\rangle\langle\phi| + \braket{\phi|\psi}\ket{\phi}\bra{\psi} \ge 0
\quad \Leftrightarrow \quad \braket{\psi|\phi}=0,1.
\end{gather}
That is, $\Oc$ is a positive operator if and only if $\ket{\psi}$ and $\ket{\phi}$ are orthogonal to each other. Here $\ket{\psi}$ and $\ket{\phi}$ are vectors in $d$--dimensional Hilbert space~(\cite{arXiv:1201.1212}).

For concreteness let us suppose that a 2--level system, a qubit, can be prepared into any possible pure state. It is easy to show that states with orthogonal vectors in the Bloch sphere lead to the maximum violation of classicality. Take for instance the anticommutator of states $\rho_X=\ket{+}\bra{+}$ with $ \ket{+}=(\ket{0}+\ket{1})/\sqrt{2}$ and $\rho_Z=\ket{0}\bra{0}$:
\begin{gather}
\Oc=\frac{1}{\sqrt{2}} \left(\ket{+}\bra{0} + \ket{0}\bra{+} \right),
\end{gather}
which has eigenvalues $\lambda_{\pm}=1 \pm \sqrt{2}$.

%%%%%%%%%%%%%%%%%%%%%%%%%%%%%%%%%%%%%%%%%%%%%%%%%%%%%%%%%%%%%%%
\subsection{Mixed states}
%%%%%%%%%%%%%%%%%%%%%%%%%%%%%%%%%%%%%%%%%%%%%%%%%%%%%%%%%%%%%%%

For the rest of the article we will only work with qubits for simplicity. However, all claims that we will make can be generalized to $d$--dimensional systems in a straightforward manner.

A mixed state of a qubit can always be written as a mixture of a pure state and a fully mixed state. Consider the two states
\begin{gather}\label{ms}
\rho_X= \frac{1-p}{2}\mathbb{I} + p \ketbra{+} \quad {\rm and} \quad
\rho_Z= \frac{1-p}{2}\mathbb{I} + p \ketbra{0}.
\end{gather} 
The anticommutator of these two states is not positive-definite only for
\begin{gather}
p > p_c = \frac{1}{\sqrt{2}},
\label{pc}
\end{gather}
where $p_c$ denotes a critical value of the mixedness parameter $p$. For $p \le p_c$ we have classicality $\Oc \ge 0$ and for $p>p_c$ we have quantumness $\Oc \ngeqslant 0$.

This implies that single qubit state with less purity than $\tr[\rho^2]=3/4$ cannot be considered quantum if only the anticommutator of the states in Eq.~(\ref{ms}) is considered. This is clearly problematic and has to be dealt with somehow. For instance NMR systems are highly mixed yet still retain quantumness to some extent. We will deal with this problem in the next section.

%%%%%%%%%%%%%%%%%%%%%%%%%%%%%%%%%%%%%%%%%%%%%%%%%%%%%%%%%%%%%%%
%%%%%%%%%%%%%%%%%%%%%%%%%%%%%%%%%%%%%%%%%%%%%%%%%%%%%%%%%%%%%%%
\section{Highly mixed states of NMR}
%%%%%%%%%%%%%%%%%%%%%%%%%%%%%%%%%%%%%%%%%%%%%%%%%%%%%%%%%%%%%%%
%%%%%%%%%%%%%%%%%%%%%%%%%%%%%%%%%%%%%%%%%%%%%%%%%%%%%%%%%%%%%%%

NMR systems have very stable states that are highly controllable~(\cite{RevModPhys.76.1037}). However these states are also highly mixed; in the form of Eq.~(\ref{ms}), the value of $p_{\rm NMR} \simeq 10^{-5}$. This is much less than the critical value, i.e. $p_{\rm NMR} \ll p_c$. Our quantumness witness will not be able to detect quantumness from an NMR system by simply looking at the anticommutator of the states Eq.~(\ref{ms}). On the other hand NMR systems are ensemble-systems. That is, there are a lot of qubits available all at once, and we will make use of this fact to regain quantumness. We now explore three approaches to overcome the obstacle at hand.

%%%%%%%%%%%%%%%%%%%%%%%%%%%%%%%%%%%%%%%%%%%%%%%%%%%%%%%%%%%%%%%
\subsection{Purification approach}
%%%%%%%%%%%%%%%%%%%%%%%%%%%%%%%%%%%%%%%%%%%%%%%%%%%%%%%%%%%%%%%

Suppose we are able to apply some cooling algorithm~(\cite{Boykin}) to our highly mixed ensemble and extract a sub-ensemble of pure states. Suppose we have $n$ qubits, each in state $\rho$. The entropy of each qubit is $S(\rho)$. The total entropy of the whole ensemble is $nS(\rho)$. Using algorithmic cooling we can then extract $m=n[1-S(\rho)]$ pure qubits. For $p_\textrm{NMR}=10^{-5}$ and $n=10^{23}$ the value of $m=10^{12}$. The quantumness of a system with this many qubits should be very high, assuming we can simultaneously prepare them all in two complementary directions. In fact, we only need to purify a hand full of qubits in order to test our witness. On one hand this is a nice solution, but may not be satisfying as this method bypasses the central problem, that our anticommutator witness $\Oc$ in Eq.~(\ref{witness1}) does not perceive the quantumness of most mixed states. Next we explore how we can test the witness without applying a purification procedure.

%%%%%%%%%%%%%%%%%%%%%%%%%%%%%%%%%%%%%%%%%%%%%%%%%%%%%%%%%%%%%%%
\subsection{Quantumness in large numbers}
%%%%%%%%%%%%%%%%%%%%%%%%%%%%%%%%%%%%%%%%%%%%%%%%%%%%%%%%%%%%%%%

Note that if we input more than one copy of the input states, the behavior (i.e.\ the positivity) of the witness may change. Consider modifying the witness in Eq.~(\ref{witness1}) as
\begin{gather}\label{witnessN}
\Oc^{(n)}= \left\{ \rho^{\otimes n}_X , \rho^{\otimes n}_Z \right\} = \rho^{\otimes n}_X \rho^{\otimes n}_Z + \rho^{\otimes n}_Z \rho^{\otimes n}_X.
\end{gather}
Individually, each qubit can be prepared in two possible states $\rho_X$ and $ \rho_Z$ as in Eq.~(\ref{ms}): in both cases, the states can be considered highly ``classical", in the sense that $p$ is well below the critical value given in Eq.~(\ref{pc}). Moreover, a single-particle anticommutator witness like in Eq.~(\ref{witness1}) would not be able to detect any quantumness. However, the witness in Eq.~(\ref{witnessN}) considers the quantumness of $n$ qubits at the same time. The ensemble of qubits has amplifie-coherence and therefore the ensemble of qubits ought to be considered a single system. After all it is the ensemble average that is measured in NMR systems.

\begin{figure}[t]
\begin{center}
\resizebox{7.2 cm}{4.74 cm}{\includegraphics{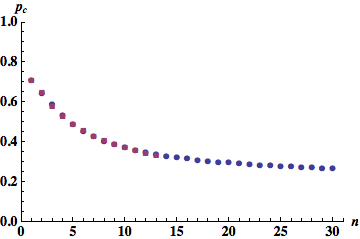}}
\end{center}
\caption{\label{plot1} Critical values of mixedness parameter for $\Oc^{(n)}$. Dots are the computed critical values of the mixedness parameter, and squares are the values of the fitted curve from Eq.~(\ref{fit1}). The critical value is never less than $p_\textrm{NMR} \simeq 10^{-5}$.}
\end{figure}

Indeed we can easily verify that the critical value of the mixedness parameter decreases simply by introducing extra qubits. For one qubit the critical value is $p_c=1/\sqrt{2}$ as given in Eq.~(\ref{pc}), for two qubits $p^{(2)}_c=0.644$, for three qubits $p^{(3)}_c=0.578$, and so on. We have computed the critical value of the mixedness parameter for $n=1, \dots, 13$. Computing the eigenvalues of $\Oc^{(n)}$ for larger $n$  becomes very difficult since the size of the matrices grows exponentially. However we have fitted the numerical data to a curve and found that it matches very well with a polynomial function
\begin{align}\label{fit1}
&  p^{(n)}_c=A_0 + \frac{A_1}{n}  + \frac{A_2}{n^2}  + \frac{A_3}{n^3}, \\
\rm{with} \quad &  A_0 = 0.204, \; A_1 = 1.882, \; A_2 = -2.660, \; A_3 = 1.281. \nonumber
\end{align}
(Notice that fits with exponential test functions do not yield good results.)
Both the numerical results and the fitted curve are plotted in Fig.~\ref{plot1}. 
Remarkably, as $n \to \infty$ the mixedness parameter has a finite critical value $p_c^{(n)} \to A_0\gg p_\textrm{NMR}$. This implies in an NMR system, without any preprocessing, one cannot witness quantumness using the witness in Eq.~(\ref{witnessN}). We have once again failed to detect quantumness in an NMR system. We now allow some preprocessing to see if that helps the emergence of quantumness.
%%%%%%%%%%%%%%%%%%%%%%%%
%%%%%%%%%%%%%%%%%%%%%%%%

%CHANGED HERE
We have once again failed in detecting quantumness in an NMR system. We now allow some preprocessing to see if that helps the emergence of quantumness.

%%%%%%%%%%%%%%%%%%%%%%%%%%%%%%%%%%%%%%%%%%%%%%%%%%%%%%%%%%%%%%%
\subsection{Quantumness in large numbers and correlations}
%%%%%%%%%%%%%%%%%%%%%%%%%%%%%%%%%%%%%%%%%%%%%%%%%%%%%%%%%%%%%%%

One of the strong points of NMR systems is the unitary gates: they are easy to implement and have high fidelity. We can use the unitary gates to build correlations between qubits and perhaps the witness will be sensitive to the correlations. In the light of that, let us once again redefine our witness operators given in Eq.~(\ref{witnessN}) as
\begin{gather}\label{witnessNC}
\Oc^{(n)}_\mathcal{C}= \left\{ U \rho^{\otimes n}_X U^\dag, V \rho^{\otimes n}_Z V^\dag \right\} = U \rho^{\otimes n}_X U^\dag V \rho^{\otimes n}_Z V^\dag + V \rho^{\otimes n}_Z V^\dag U \rho^{\otimes n}_X U^\dag,
\end{gather}
where $U$ and $V$ are unitary operations that build correlations in $\rho_X$ and $\rho_Z$, respectively. There is great deal of freedom in choosing $U$ and $V$. In fact, optimizing the witness over all possible encodings does not seem trivial and we will leave to later studies on this problem.

We have studied one such case: we implement controlled-\textsc{not} gates, using the first qubit as control and the rest as target. The controlled-\textsc{not} gates are applied in the appropriate basis to achieve maximal correlations in the final state. Specifically, the gates are
\begin{align}
\label{UV}
U &= \ketbra{0} \otimes \mathbb{I}^{\otimes(n-1)} + \ketbra{1} \otimes \sigma_X^{\otimes(n-1)},\\
V &= \ketbra{+} \otimes \mathbb{I}^{\otimes(n-1)} + \ketbra{-} \otimes \sigma_Z^{\otimes(n-1)},
\end{align}
where $\sigma_{X}$ and $\sigma_Z$ are Pauli operators. We should note the correlations in the final state are purely classical. Nevertheless the correlations enhance the quantumness of the states as measured by the witness in the last equation. 

\begin{figure}[t]
\begin{center}
\resizebox{7.2 cm}{4.74 cm}{\includegraphics{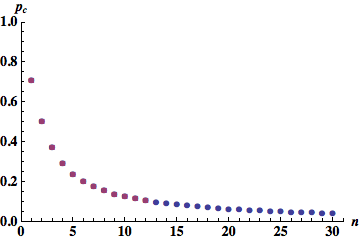}}
\end{center}
\caption{\label{plot2} Critical values of mixedness parameter for $\Oc^{(n)}_\mathcal{C}$. Dots are the computed values of the critical mixing point, and squares are the values of the fitted curve Eq.~(\ref{fit2}). For $n\simeq 140,000$ the critical value is expected to be less than $p_\textrm{NMR} \simeq 10^{-5}$.}
\end{figure}

Once again we have numerically computed the critical value of the mixedness parameter for the preprocessed states. We numerically computed the critical values of the mixedness parameter for $n = 1, \ldots, 12$ and plotted the numerical data, along with the fitted curve, in Fig.~\ref{plot2}. The fitted curve is once again polynomial of the form
\begin{align}
\label{fit2}
& p^{(n)}_c = B_0 + \frac{B_1}{n}  + \frac{B_2}{n^2}  + \frac{B_3}{n^3} , \\
\rm{with} \quad & B_0 = -0.006, \; B_1= 1.350, \; B_2 = -0.711, \; B_3 = 0.075. \nonumber
\end{align}
(We stress that fits with exponential test functions do not yield good results and that adding more terms do not yield better results, in that $B_3$ is sensibly smaller than the other coefficients.) Note that the leading coefficient is compatible with zero, implying that as $n \to \infty$ the critical value of the mixedness parameter $p_c^{(n)} \to 0$. We can therefore set $B_0 = 0$ and compute the number of qubits needed to achieve $p_c^{(n)} < p_\textrm{NMR}$. This turns out to be around $n_\textrm{NMR}\simeq 1.35 \times 10^5$.
%%%%%%%%%%%%%%%%%%%%%%%%
%%%%%%%%%%%%%%%%%%%%%%%%

An analogue to the witnesses in Eqs.~(\ref{witnessN}) and (\ref{witnessNC}) using the observable approach of AvR is difficult. For the observable approach it would simply mean that you would define some observable in the space of $n$ qubits and look for states of the system that would lead to quantumness. Here we have to be able to keep the state of the system fixed, but simply change the number of systems that lead to quantumness. This gives an operational meaning to mixed state ensembles. It is worth noting that quantumness arises in the large number limit. For the example above we would need $\simeq 1.35 \times 10^5$ qubits to witness it. This is contrary to the popular notion that quantumness goes to classicality in the large number limit.

On the other hand witnessing this quantumness is not trivial. For the experimental prescription we laid out in section \ref{OQ}, we would need $1.35 \times 10^5$ sets of $1.35 \times 10^5$ qubits. Then we would have to implement a \textsc{shift} operator among all of these qubits. That is, we would need over ten billion qubits and be able to interact them all. This is far from what is achievable with the current technology. We would need to reduce the number of qubits necessary somehow in order to witness this sort of quantumness in NMR, perhaps using some different preprocessing techniques.

%%%%%%%%%%%%%%%%%%%%%%%%%%%%%%%%%%%%%%%%%%%%%%%%%%%%%%%%%%%%%%%
%%%%%%%%%%%%%%%%%%%%%%%%%%%%%%%%%%%%%%%%%%%%%%%%%%%%%%%%%%%%%%%
\section{Conclusion}
%%%%%%%%%%%%%%%%%%%%%%%%%%%%%%%%%%%%%%%%%%%%%%%%%%%%%%%%%%%%%%%
%%%%%%%%%%%%%%%%%%%%%%%%%%%%%%%%%%%%%%%%%%%%%%%%%%%%%%%%%%%%%%%

We have extended the work of~(\cite{AVR, APVR}) and have given a quantumness witness in terms of states of the system. Our witness works in the same spirit as the work of AvR, but we find some surprising results. Namely that the witness does not detect quantumness for even moderately mixed states of the system. We overcame this obstacle by noting that the witness sees more quantumness as the number of the system state grows. This gives us an operational interpretation of quantumness for ensemble size. Further, our quantumness witness allows us to deal with correlations in a natural way.

Though quantumness grows with the number of systems, detecting the witness remains difficult. This is because when working with $n$ qubits, we need $n$ sets of $n$ qubits. If $n$ were a marginally large number, we would need to work with a reasonable number of qubits. Otherwise this test is not possible with current technology. Perhaps different correlation class of states will lower this number significantly and allow for observation quantumness in NMR-like systems.

Lastly, we note our witness is related to quantum discord~(\cite{arXiv:1201.1212}) using the necessary and sufficient criteria for classicality of correlations given in~(\cite{arXiv:1005.4348}). This is interesting due to the fact that when classically correlated states are put together, a quantum correlated state emerges within the discord framework of quantumness of correlations. This begs the question: should we consider quantum mechanics to be the basis for classical physics? Perhaps the universe if a bit more complicated than that.

{\bf Acknowledgements.} KM and VV thank the John Templeton Foundation for support. KM, RF, and VV acknowledge the financial support by the National Research Foundation and the Ministry of Education of Singapore. SP and KY would like to thank the Centre for Quantum Technologies at the National University of Singapore for the hospitality. SP and KY are partially supported by the Joint Italian-Japanese Laboratory on ``Quantum Technologies: Information, Communication and Computation'' of the Italian Ministry for Foreign Affairs. KY is supported by the Ministry of Education, Culture, Sports, Science and Technology, Japan.

\bibliographystyle{rspublicnat}
\bibliography{NMRSI}
\end{document}